\documentclass[12pt]{article}

\usepackage{graphicx}

\begin{document}

\begin{center}
{\large{\bf {On 2D Periodic Hexagonal Cells}}
\vspace{0.5in}

\noindent{\small{\em R. Delbourgo\footnote{electronic address: 
 \tt {\bf bob.delbourgo@utas.edu.au}}}}
\medskip

{\small \noindent School of Mathematics and Physics, 
        University of Tasmania\\
        GPO Private Bag 37, Hobart, Australia 7001}}
\vspace{1in}
\end{center}

\date{\today}

\begin{abstract}
Graphene, the new wondrous material, is a perfect example of a two-dimensional
hexagonal crystal unlike any other. Here we exhibit some of the 
characteristic directional features associated with hexagonal cells, emphasising 
the sixfold symmetry. We depict the X-ray, vibrational and electronic band structures 
to be expected in such systems via 2 dimensional contour plots.
\end{abstract}
 

\newpage

\section{Introduction} 

It is little wonder that graphene has captured the interest of physicists or materials scientists and 
no surprise that it is the centrepiece of the 2010 Nobel Physics Prize. The reason for the great
excitement lies in is its remarkable transport properties and amazing strength plus the
fact that technology has advanced to the stage where graphene can be produced in relatively
large sheets, on a variety of substrates. A very bright future is predicted for it. Thus it possesses
very low resistivity, high opacity and can furnish the photonics, spintronics and electronic industry
with a low cost alternative to silicon.

There are numerous review articles \cite{GN07,GK08,G09,CGPNG09,ATK10} 
on this subject and a veritable avalanche of research papers
has appeared, describing or exploring different properties. These concern its electronic characteristics, its vibrational properties, its edge effects, the Dirac-like properties of quasiparticles 
at the corners of Brillouin zones \cite{McClure, Charlier,Netal05,MRSR10}, anomalous Hall effects, etc.  
Our aim in this paper is very modest: 
we wish to illustrate its beautiful directional properties in a manner that might benefit teachers of
condensed matter physics: the 2D hexagonal symmetry is somewhat unfamiliar and can
serve as a nice extension of the standard square symmetry analysis \cite{Wallace}.

The paper is set out as follows. Section 2 contains the basic notation and its consequences
for glancing X-rays. In Section 3 we discuss the oscillation modes within a
hexagon, clamped at it edges. These represent the energy levels of a particle held in
the infinite hexagonal well; the results are no doubt familiar to the graphene experts
but are probably not widely known to the usual quantum mechanics practitioners, 
who often resort to rectangular boundary conditions. We also consider the modes 
corresponding to antinodes at the edges because these are subsequently
needed for discussing the Kronig-Penney model in Section 5. Section 4 is devoted to
idealised vibrational modes (longitudinal and transverse) of the atoms themselves, 
assuming nearest neighbour interactions. Here, as in Section 3, our aim is to illustrate
the hexagonal symmetry of the motions. In the last Section we describe the electronic
band structure, using the approximation of delta-function walls between adjoining cells
rather than the tight-binding approximation used in most expositions. In order to keep
the analysis simple, nowhere do we study edge effects (armchair, zigzag, corrugations 
\cite{MP09}) or multilayer graphene; for such elaboration we refer the reader to research 
reviews. We hope that this paper will serve as a useful extension of the usual quantum 
mechanical problems encountered by college students and highlight the beautiful
hexagonal symmetry of the principal properties of graphene.

\section{Bases and X-Ray Features} 

Let us set down our notation which will be used in subsequent sections. The basic
hexagonal cell (with side length $\ell$) is drawn in Figure 1 and our coordinate system
is centred there. The fundamental cell vectors are
\begin{equation}
 {\bf{a}} = (3,-\sqrt{3})\,\ell/2, \qquad{\bf{b}}=(3,\sqrt{3})\,\ell/2.
\end{equation}
Any two nearby corner points such as carbon atoms at U and V can be taken
as centres of X-ray scattering so that all lattice points are generated by translation:
\begin{eqnarray}
{\bf{u}}_{rs}={\bf{u}}_{00}+r{\bf{a}}+s{\bf{b}}&=&(3r+3s-1,\sqrt{3}(s-r-1))\ell/2 \nonumber \\
{\bf{v}}_{rs}={\bf{v}}_{00}+r{\bf{a}}+s{\bf{b}}&=&(3r+3s+1,\sqrt{3}(s-r-1))\ell/2 ,
\end{eqnarray}
where $r$ and $s$ are integers. 
The basic reciprocal lattice, drawn in Figure 2, is nothing more than a scaled version
of the original lattice, rotated by $30^o$ with side length $4\pi/3\sqrt{3}\ell$.
Its fundamental vectors are
\begin{equation}
 {\bf{A}} = (1,-\sqrt{3})\,2\pi/3\ell, \qquad{\bf{B}}=(1,\sqrt{3})\,2\pi/3\ell.
\end{equation}
Then, according to the Von Laue conditions for constructive interference, whenever
the change of wave number is ($R$ and $S$ are integers)
\begin{equation}
\Delta{\bf{k}} = R{\bf{A}}+S{\bf{B}}=(R+S,(S-R)\sqrt{3})\,2\pi/3\ell,
\end{equation}
X-rays directed {\em along} the graphene plane will of course be diffracted by the lattice. However
because there are {\em two} atoms decorating each unit cell there is a modulating factor
$1 + \exp[(R+S)\,2\pi/3] \propto 2\,\cos[\pi(R+S)/3]$ apart from an irrelevant phase; this means 
that diffraction will be greatest when $R+S$ is exactly divisible by 3 but that there are no missing 
orders of diffraction.

\section{The hexagonal well} 

The literature on the equilateral triangular drum is old indeed; it goes back to Lam\'{e} 
\cite{Lame} and Pockels \cite{Pockels}. A very nice series of reviews on this topic has been 
written by McCartin \cite{McCartin}, discussing the completeness, orthogonality  and multiplicity
of the solutions, subject to Dirichlet and Neumann conditions at the edge of the triangle. 
It is not hard to adapt the solutions to the case of the hexagon and we shall do so shortly. 
(As a matter of fact the solutions of the equilateral triangle are embedded in a one-sixth 
section of the hexagon.) Here is an outline of the main steps.

\begin{figure}[htbp] 
   \centering
   \includegraphics[width=2in]{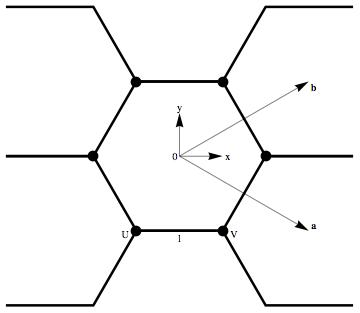} 
   \caption{Basic cell vectors. Any two adjoining C atoms such as U 
     and V can be taken as the basic cell components.}
   \label{Fig1Label}
\end{figure}

First of all introduce triangular coordinates, $u,v,w$ centred at the origin with $u+v+w=0$, 
as illustrated in Figure 3. The maximum values of these coordinates within the hexagon is
the radius of the incircle $r=\ell\sqrt{3}/2$ and we see that in terms of the normal Cartesian
coordinates,
\begin{equation}
 u = -y, \qquad  v-w=\sqrt{3}\,x.
\end{equation}

\begin{figure}
  \centering
   \includegraphics[width=2in]{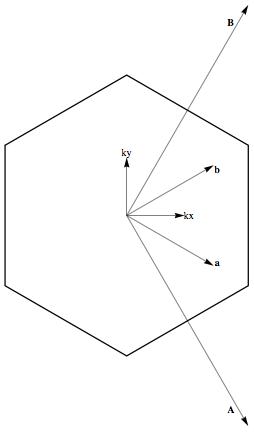} 
   \caption{Brillouin zone and its basis vectors \bf{A} and \bf{B}.}
   \label{Fig2Label}
\end{figure}

\begin{figure}
\centering
   \includegraphics[width=2in]{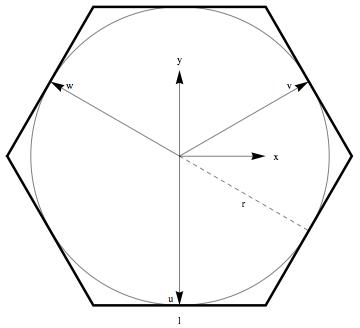} 
   \caption{Triangular coordinates for describing modes of oscillation.}
   \label{Fig3Label}
\end{figure}

The second step is to note that the Schr\"{o}dinger or Helmholtz equation for a free 
particle of energy $E$ in the cell, namely
\[
\Big[ \frac{\partial^2}{\partial x^2} + \frac{\partial^2}{\partial y^2} + k^2 \Big] \psi(x,y) =0;
\quad k^2=\frac{2ME}{\hbar^2},
\]
may be transcribed into triangular coordinates:
\begin{equation}
\Big [ \frac{\partial^2}{\partial u^2} + 3\frac{\partial^2}{\partial (v-w)^2} + k^2 \Big] \psi(x,y) =0,
\end{equation}
cleverly allowing for cyclic sums of separable solutions of the form,
\[
\psi(u,v,w)=[\cos(k_1u)\!+\!C\sin(k_1u)]\,
                [\cos(k_2\!-\!k_3)\frac{(v\!-\!w)}{3}+\!D\sin(k_2\!-\!k_3)\frac{(v\!-\!w)}{3}]
 \]
 \[
 \quad+\, [\cos(k_2u)+C\sin(k_2u)]\,
               [\cos(k_3\!-\!k_1)\frac{(v\!-\!w)}{3}+\!D\sin(k_3\!-\!k_1)\frac{(v\!-\!w)}{3}]
 \]
 \begin{equation}              
 \qquad\,\,\,+ \,[\cos(k_3u)+C\sin(k_3u)]\,
                [\cos(k_1\!-\!k_2)\frac{(v\!-\!w)}{3}+\!D\sin(k_1\!-\!k_2)\frac{(v\!-\!w)}{3}] 
\end{equation}
where $k_1+k_2+k_3 = 0$ and $k^2=2(k_1^2 + k_2^2+k_3^2)/3 = 4(k_1^2+k_2^2+k_1k_2)/3$. 
The third step is to pick out symmetric or antisymmetric solutions under reflection about
a vertical diagonal; this selects out $\sin$ or $\cos$ functions in fact. Typically one finds \cite{McCartin}
the (Dirichlet) quantized solutions $k_i=\pi n_i/r = 2\pi n_i/\sqrt{3}\ell$ -- the case of 
interest for this section:
\begin{eqnarray}
 \psi_{D\,sym/anti} &=& \sin\frac{\pi n_1u}{r}.\cos/\sin\frac{\pi(n_2-n_3)(v-w)}{3r} + \nonumber\\
 & & \sin\frac{\pi n_2u}{r}.\cos/\sin\frac{\pi(n_3-n_1)(v-w)}{3r} + \nonumber\\
 & & \sin\frac{\pi n_3u}{r}.\cos/\sin\frac{\pi(n_1-n_2)(v-w)}{3r}\,\, ,
\end{eqnarray}
where $n_i$ are non-zero integers subject to $n_1+n_2+n_3=0$. It suffices to
take $n_1\geq n_2>0$ to obtain distinct modes (in the antisymmetric configuration 
$n_1$ and $n_2$ need to be different for a nonvanishing answer). Hence the
quantized energy levels are given by 
\begin{equation}
k_{n1,n2,n3} ^2 = 8\pi^2(n_1^2+n_2^2+n_3^2)/9\ell^2 = 16\pi^2(n_1^2+n_2^2+n_1n_2)/9\ell^2.
\end{equation}

While we are discussing this topic let us note \cite{McCartin} the Neumann solutions (effectively antinodes along the hexagon edges) as we will require them later:
\begin{eqnarray}
 \psi_{N\,sym/anti} &=& \cos\frac{\pi n_1u}{r}.\cos/\sin\frac{\pi(n_2-n_3)(v-w)}{3r} + \nonumber\\
 & & \cos\frac{\pi n_2u}{r}.\cos/\sin\frac{\pi(n_3-n_1)(v-w)}{3r} + \nonumber\\
 & & \cos\frac{\pi n_3u}{r}.\cos/\sin\frac{\pi(n_1-n_2)(v-w)}{3r}\,\, .
\end{eqnarray}

We have illustrated some of the lowest few modes in Figures 4 to 6 in the form of contour plots rather
than 3D plots as they are easier to comprehend (the lightest shaded ares are the maxima while
the darkest correspond to minima). They show the lovely hexagonal features expected of
the vibrations. The fundamental Neumann mode is depicted in Figure 7 even though it is
not required for the present purposes. [It should be noted that triangular cuts within the hexagon describe the modes for an equilateral triangle.] McCartin has analysed the orthogonality and 
completeness properties of solutions (8) and (10) so we refer the reader to his review article.

\begin{figure}
   \centering
   \includegraphics[width=3in]{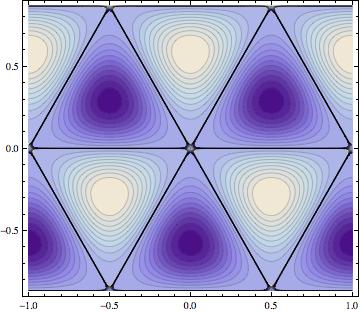} 
   \caption{Fundamental mode (1,1) of the unit hexagonal well.}
   \label{Fig4Label}
\end{figure}

\begin{figure}
   \centering
   \includegraphics[width=3in]{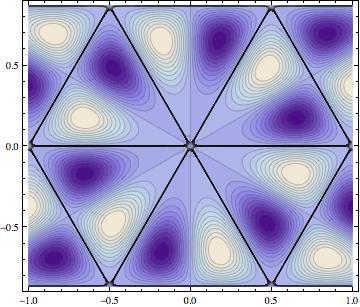} 
   \caption{Antisymmetrical overtone (2,1) of the unit hexagonal well.}
   \label{Fig5Label}
\end{figure}

\begin{figure}
 \centering
   \includegraphics[width=3in]{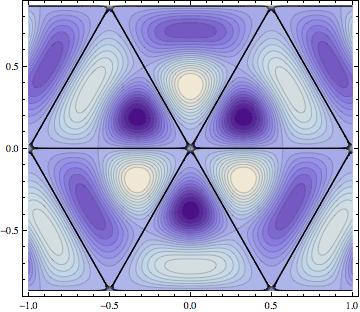} 
   \caption{Symmetrical overtone (2,1) of the unit hexagonal well.}
   \label{Fig6Label}
\end{figure}

\begin{figure}
   \centering
   \includegraphics[width=3in]{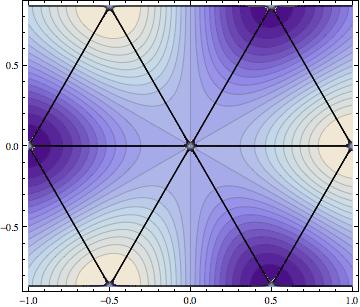} 
   \caption{Basic Neumann mode (1,0) of the unit hexagonal well.}
   \label{Fig7Label}
\end{figure}

\newpage
\section{Vibrational Features} 

Let us begin by tackling vertical (out of plane) transverse oscillations. Regarding the graphene
sheet as some sort of trampoline under fixed tension and considering only nearest
neighbour interactions, the downward acceleration on a C atom located at ${\bf u}_{rs}$ is 
determined by the adjoining vertical displacements relative to the atoms at 
${\bf v}_{rs}, {\bf v}_{r s-1}, {\bf v}_{r-1\,s}$. Hence vibrations are described by the simple 
harmonic equation,
\begin{equation}
  d^2{U}_{rs}/dt^2 = -\Omega^2(3U_{rs} - V_{rs} - V_{r s-1}-V_{r-1\,s}),
\end{equation}
where $\Omega$ is a characteristic frequency associated with the ``tension'' of the links and
how strongly or loosely the C atoms adhere to the substrate. Similarly for the twin atom located
at  ${\bf v}_{rs}$ one obtains
\begin{equation}
 d^2{V}_{rs}/dt^2 = -\Omega^2(3V_{rs} - U_{rs} - U_{r s+1}-U_{r+1\,s}).
\end{equation}
Looking for plane wave solutions [${\bf k}=(k_x,k_y)$ below], at the positions summarized by (2),
\[
U_{rs} = U\exp[i({\bf k}.{\bf u}_{rs})-\omega t],\quad V_{rs} = V\exp[i({\bf k}.{\bf v}_{rs})-\omega t]
\]
one arrives at two coupled equations for the amplitudes of the cell atoms  at U and V:
\begin{eqnarray}
 \omega^2\,U&=&\Omega^2[3\,U - V{\rm e}^{ik_x\ell} - 2V{\rm e}^{-ik_x\ell/2}\cos(k_y\ell\sqrt{3}/2)]
       \nonumber \\
 \omega^2\,V&=&\Omega^2[3\,V - U{\rm e}^{-ik_x\ell} - 2U{\rm e}^{ik_x\ell/2}\cos(k_y\ell\sqrt{3}/2)],
\end{eqnarray}
producing the eigenvalue equation
\[
(\omega^2/3\Omega^2-1)^2 = (1+4c_y^2+4c_xc_y)/9;\quad c_x\equiv \cos(3k_x\ell/2),
   c_y\equiv\cos(\sqrt{3}k_y\ell/2)
\]
or dispersion relation between frequency and wavenumber for transverse vertical acoustic (-) 
and optical (+) modes:
\begin{equation}
\omega_{\pm}=\Omega\,[\,3 \pm \sqrt{1+4c_xc_y+4c_y^2}].
\end{equation}
This dispersion relation for the two modes is shown in Figure 8. In the long wavelength limit,
\begin{equation}
 \omega_-^2/\Omega^2 \simeq  3\ell^2(k_x^2+k_y^2)/4, \quad {\rm so} \quad 
  v_{acoustic} =2\Omega/\sqrt{3}\ell,
\end{equation}
whereas $\omega_+ \simeq \sqrt{6}\Omega$ for optical vibrations. The acoustic and optical
modes remain separated when the wave vector $\bf k$ is directed along $x$ even at the edge
of the Brillouin zone, but they merge when $\bf k$ is directed along $y$ and a corner (a so-called
Dirac point) of the zone is approached. It is the instabilities in the vertical motion that apparently
tend to make graphene sheets ripple or even curl up into tubes.

\begin{figure}
   \centering
   \includegraphics[width=4in]{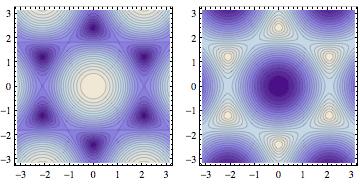} 
   \caption{$\omega-k$ dispersion relation for acoustic and optical transverse
     modes (scaled by $\Omega$) within the fundamental Brillouin zone.}
   \label{Fig8Label}
\end{figure}

Next we shall study the oscillations of the C atoms within the graphene plane. This time we
are dealing with two-dimensional vector displacements ${\bf U} =(U_x,U_y),\,{\bf V}=(V_x,V_y)$
for the atoms located at positions determined by (2), rather than the vertical displacements
considered previously. For small vibrations we assume that the restoring force is proportional
to the extension in the direction of the link; so if one defines three unit vectors ${\bf e}_{1,2,3}$ along
each of the links, the total restoring force of the atom at U summed over the three links is
\[
[ ({\bf V}_{rs}-{\bf U}_{rs}).{\bf e}_1]{\bf e}_1 + [ ({\bf V}_{r-1\,s}-{\bf U}_{rs}).{\bf e}_2] {\bf e}_2
+ [ ({\bf V}_{rs-1}-{\bf U}_{rs}).{\bf e}_3]{\bf e}_3
\]
apart from an overall factor. Resolving the $\bf e_i$ into Cartesian components, we get
the equation pair for the U atom ($\varpi$ is the horizontal characteristic frequency),
\begin{eqnarray}
 \!\!\!\ddot{U}_{x\,rs}/\varpi^2\!\!\!\!\!&\!=\!&
            \!\!\!\!-3U_{x\,rs}/2\! +\! V_{x\,rs}\!+\! (V_{x\,r\!-\!1\,s}\!+\!V_{x\,rs\!-\!1})/4\!+\!
            \sqrt{3}(V_{y\,r\,s\!-\!1}\! -\! V_{y\,r\!-\!1\,s})/4  \\
\!\!\!\ddot{U}_{y\,rs}/\varpi^2\!\!\!\!\!&=&
\!\!\!\! -3U_{y\,rs}/2\! +\! 3(V_{y\,r-1\,s}\!+\!V_{y\,rs-1})/4\! +\!
    \sqrt{3}(V_{x\,r\,s-1}\! -\! V_{x\,r-1\,s})/4 .   
\end{eqnarray}
A similar equation applies to the vector displacement $\bf V$ of the V atom alongside:
\begin{eqnarray}
 \!\!\!\ddot{V}_{x\,rs}/\varpi^2\!\!\!\!\!&\!=\!&
      \!\!\! -3V_{x\,rs}/2\! +\! U_{x\,rs}\!+\! (U_{x\,r+1\,s}\!+\!U_{x\,rs+1})/4\!+\!
    \sqrt{3}(U_{y\,r\,s+1}\! -\! U_{y\,r+1\,s})/4  \\
 \!\!\!\ddot{V}_{y\,rs}/\varpi^2\!\!\!\!\!&=&\!\!\!\!-3V_{y\,rs}/2\! +\! 3(V_{y\,r+1\,s}+V_{y\,rs+1})/4\! +\!
    \sqrt{3}(U_{x\,r\,s+1}\! -\! U_{x\,r+1\,s})/4 .   
\end{eqnarray}
For vibrations of circular frequency $\omega$ and wave number $\bf k$, one arrives at the
secular equation,
\begin{equation}
 \left( \begin{array}{cccc}
          W & 0 & 1+c_y{\rm e}^{-i\chi}/2 & -i\sqrt{3}s_y{\rm e}^{-i\chi}/2\\
          0 & W & -i\sqrt{3}s_y{\rm e}^{-i\chi} & 3c_y{\rm e}^{-i\chi}/2 \\
         1+c_y{\rm e}^{i\chi}/2 & i\sqrt{3}s_y{\rm e}^{i\chi}/2 & W & 0 \\
         i\sqrt{3}s_y{\rm e}^{i\chi}/2 & 3c_y{\rm e}^{i\chi}/2  & 0 & W
         \end{array} \right) 
         \left( \begin{array}{c}  
          U_x \\ U_y \\ V_x{\rm e}^{ik_x\ell} \\ V_y {\rm e}^{ik_x\ell} 
          \end{array} \right) = 0    , 
\end{equation}
where $W\equiv \omega^2/4\varpi^2 - 3/2$, $\chi \equiv 3k_x\ell/2$, 
$c_y\equiv \cos(\sqrt{3}k_y\ell/2)$, $s_y\equiv \sin(\sqrt{3}k_y\ell/2)$ and $\varpi$
stands for a characteristic frequency of planar oscillations, which is not necessarily
equal to the value $\Omega$ for vertical oscillations because of the the substrate's
influence.

This leads to the set of roots (dispersion relations):
\begin{equation}
\omega/\varpi = 0, \quad 2\sqrt{3}, \quad  (3 - \sqrt{1+4c_xc_y+4c_y^2} ),
                              \quad   (3 + \sqrt{1+4c_xc_y+4c_y^2} ).
\end{equation}
The first two roots  may be ignored (the first is static displacement and the second is
dispersionless). However, the third and fourth planar roots are strikingly similar to the 
vertical motion roots (14) (acoustic and optical), except that the oscillation frequency is not necessarily the same.

\section{Kronig-Penney electron band model}
We now turn to a model of the electronic energy bands which is a generalization of the
one-dimensional Kronig-Penney scheme -- instead of the tight-binding approximation which
is more popular. To that end, we envisage a series of $\delta$-function potential walls lying 
along the carbon links as that is where the electron clouds are at their most prominent. Any
stray negative charge carrier would encounter these repulsive barriers, except of course
at the cell corners where they would feel the attraction from the C nuclei.
Therefore as our model we may envisage loose electrons as moving freely within the cells 
until they hit a cell edge, where the wave function discontinuity across the cell boundary is proportional to the wave function there (which is itself continuous). This should mimic the
periodic properties and symmetry features of more realistic models.

Consider then the following parametrizations of the symmetric wave function $\Psi_0$
for the cell round the origin, drawn in Figure 1,  and the wave function $\Psi_1$ for the
cell below it (centred at $x=0,\,y=-\sqrt{3}\ell$). One must allow for both Dirichlet {\em and} 
Neumann boundary conditions -- the analogues of $\sin(kx)$ and $\cos(kx)$ that are 
included in the one-dimensional analysis. Here we do not consider antisymmetric
solutions, since the model merely serves as an example to illustrate the hexagonal features
of the bands -- the true nature of the bands can in fact only be revealed through experiments.
\begin{eqnarray}
\Psi_j &=& [A_j\sin(k_1u) + B_j\cos(k_1u)].\cos(k_1+2k_2)(v-w)/3 + \nonumber \\
          &  & [A_j\sin(k_2u) + B_j\cos(k_2u)].\cos(2k_1+k_2)(v-w)/3 - \nonumber \\
         &  &  [A_j\sin((k_1\!+\!k_2)u)\!-\! B_j\cos((k_1\!+\!k_2)u)].\cos(k_1\!-\!k_2)(v\!-\!w)/3, 
\end{eqnarray}
for cells $j=0$ or 1. There are the (dis)continuity conditions which we can apply at $v=w$ or $x=0$,
\begin{eqnarray}
 \Psi_1|_{u=r} - \Psi_0|_{u=r} &=&0, \nonumber \\
 (\partial \Psi_1/\partial u) \,|_{u=r} - (\partial \Psi_0/\partial u)\,|_{u=r}
   &=& (V/r)\,\Psi_0|_{u=r} ,
 \end{eqnarray}
as well as the Bloch periodicity conditions,
\begin{eqnarray}
  \Psi_1|_{u=r}& = & {\rm e}^{2iqr} .\Psi_0|_{u=-r} \nonumber \\
 (\partial \Psi_1/\partial u)\,|_{u=r} &=& {\rm e}^{2iqr} .
         (\partial \Psi_0/\partial u)\,|_{u=-r}\quad,
\end{eqnarray}
where $q$ is the (real) quasi-momentum and $V$ corresponds to the size of the $\delta$ function
barrier. We also remind the reader that the electronic energy is given by
$E=(2\hbar^2/3m^2)[k_1^2+k_2^2+k_1k_2].$

The equations (23) and (24) determine the allowed energy bands from the reality of $q$ via
the conditions' combination:
\[
(1/2 + \cos(2qr))\,[(\alpha + \beta)\cos(\alpha-\beta)-\beta\cos(2\alpha+\beta)-\alpha\cos(\alpha+2\beta)]+
\]
\[
\alpha(\cos\alpha -\cos 2\alpha)+\beta(\cos\beta -\cos 2\beta)-(\alpha+\beta)[\cos(\alpha+\beta)-
  \cos(2\alpha+2\beta)]
\]
\[
+\,\,V[2\sin\alpha-\sin 2\alpha+2\sin\beta-\sin 2\beta-2\sin(\alpha+\beta)+\sin(2\alpha+2\beta)]=0,
\]
where $\alpha\equiv k_1r,\beta\equiv k_2r$. With $k_1$ and $k_2$ inclined at 120$^o$ with
respect to one another and $-1\leq \cos(qr)\leq 1$, one may plot the allowed regions of $k_i$
space for the energy bands. These are drawn in Figures 9 and 10, where we see that if we follow
the $k_x$ axis, the bands get narrower as $V$ increases \cite{f1}. Indeed as $V\rightarrow\infty$
the ratio $B/A \rightarrow 0$ in equation (22) and the $k$ values become quantized as in 
the infinite hexagonal well. More generally the energy bands lie between concentric circles at low energies because of the nature of the expression $k_1^2+k_2^2+k_3^2$ with triangular
axes. For the tight binding model we refer the reader to reference \cite{Wallace}.

\begin{figure}[h]
   \centering
   \includegraphics[width=3in]{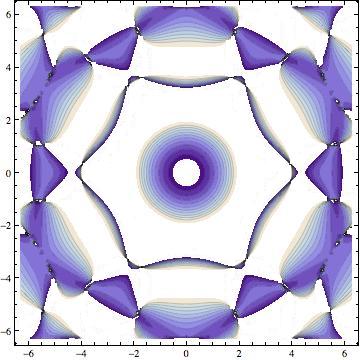} 
   \caption{Allowed band (shaded region) for $V=1$.}
   \label{Fig9Label}
\end{figure}

\begin{figure}[h]
   \centering
   \includegraphics[width=3in]{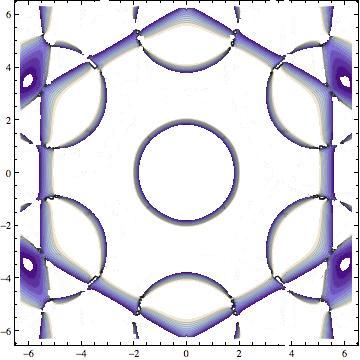} 
   \caption{Allowed band (shaded region) for $V=10$.}
   \label{Fig10Label}
\end{figure}

\section{Conclusions } 

Our aim in this paper has not been to provide an exhaustive analysis of all properties of graphene,
not even in summarized form. Plentiful reviews exist which do just that. Rather our
purpose has been to highlight the {\em symmetry} features to expect from this remarkable material
both in the electronic and vibrational characteristics, in the idealized situation that one has an
infinite lattice of it. In that way, the directional properties of graphene come much more into
evidence. In practice of course we expect the symmetry to be ruined by boundary effects,
though the departure is probably significant only near the zigzag and armchair edges.

Graphene is a perfect vehicle for teaching two-dimensional aspects of solid state physics and
makes for a very useful variation on the standard theme of square lattices. University teachers
can use the material to illustrate many aspects of condensed matter and enlarge the students'
horizons.

\section*{Acknowledgments}
The use of Mathematica has greatly eased the task of producing the various diagrams.


\begin{thebibliography}{99}

\bibitem{GN07} A.K.~Geim and K.S.~Novoselov, Nature Materials  {\bf 6}, 183 (2007).
\bibitem{GK08} A.K.~Geim and P.~Kim, Scientific American, 90, April (2008).
\bibitem{G09} A.K.~Geim, Science {\bf 324}, 1530 (2009).
\bibitem{CGPNG09} A.H.~Castro-Neto, F.~Guinea, N.M.R.~Peres, K.S.~Novoselov and A.K.~Geim,
 {Rev.Mod. Phys.} {\bf 81}, 109  (2009).
\bibitem{ATK10} M.J.~Allen, V.C.~Tung and R.B.~Kaner, Chem. Rev. {\bf 110}, 132 (2010).
\bibitem{McClure} J.W.~McClure, Phys Rev. {\bf 108}, 612 (1957); {\em ibid} Phys. Rev. {\bf 112}, 715
(1958).
\bibitem{Charlier}J.C.~Charlier, P.C.~Eklund, J.~Zhu and A.C.~Ferrari, Top. Appl. Phys. {\bf 111}
673 (2008).
\bibitem{Netal05} K.S.~Novoselov {\em et al},  Nature {\bf 438}, 197 (2005).
\bibitem{MRSR10} G.~Murguia, A.~Raya, A.~Sanchez and E.~Reyes, Am. J. Phys. {\bf 78},700 (2010).
\bibitem{Wallace} P.R.~Wallace,  {Phys. Rev.} {\bf 71}, 622 (1947).
\bibitem{MP09} A.~Matulis and F.M.~Peeters, Am. J. Phys. {\bf 77}, 595 (2009).
\bibitem {Lame} G.~Lam\'{e}, J. de l'Ecole Poly. {\bf 22}, 194 (1833).
\bibitem{Pockels} F.~Pockels, ``Uber die partielle Differentialgleichung $\Delta u + k^2 u =0$'',
B.G.~Teubner, Leipzig (1891).
\bibitem{McCartin} B.J.~McCartin, Amer. Conf. on Applied Mathematics (MATH '08) Harvard {\bf 195},
March (2008), and references therein.
\bibitem{f1} The result above may be contrasted with the well-known 1D result: $\cos\,q\ell= \cos\,k\ell +
V\,(\sin\,k\ell)/k\ell.$
\end{thebibliography}
\end{document}